\documentclass[aps,twocolumn]{revtex4}

\begin{document}

\bibliographystyle{apsrev}

\title{Rapid LISA Astronomy}
\author{ Neil J. Cornish}
\affiliation{Department of Physics, Montana State University, Bozeman, MT 59717}

\begin{abstract}
A simple method is presented for removing the amplitude, frequency and phase
modulations from the Laser Interferometer Space Antenna (LISA) data stream for sources at any
sky location. When combined with an excess power trigger or the fast chirp transform, the
total demodulation procedure allows the majority of LISA sources to be identified without recourse to
matched filtering.
\end{abstract}
\pacs{}

\maketitle

A revolution is underway that will transform the field of astronomy. An array of
ground and space based gravitational wave detectors are in various stages of design,
construction and operation. The first space based detector, the Laser Interferometer
Space Antenna~\cite{lppa}, is being developed for a launch in 2011. These gravitational
wave detectors will open a new window on the Universe that complements
traditional forms of astronomy~\cite{hughes}.

However, before we can talk about gravitational wave astronomy, a method must be found
to turn a gravitational wave detector into a gravitational wave telescope - that is,
a device that is capable of locating individual sources on the sky and determining
their physical characteristics. In traditional electromagnetic astronomy
the conversion of a detector - such as the human eye or a photographic
plate - into an observational instrument is achieved by building a device that
collects and focuses the incident radiation onto the detector. Such an approach is
impractical for gravitational radiation as the weak coupling of the radiation
to matter prevents the construction of a physical telescope. Instead, a gravitational
wave detector is transformed into a gravitational wave telescope by data analysis
algorithms.

In principle we would like to use matched filtering of the data against suitable waveform
templates as this delivers the highest fidelity results, but in practice the computational cost
of matched filtering can be prohibitive. The solution is to adopt a hierarchical approach
where a fast, but sub-optimal, method is used to find a good initial solution, which can
then be refined using matched filtering. The excess power statistic~\cite{lads} used in
ground based gravitational wave astronomy is a good example of this approach. However,
we cannot simply take the ground based data analysis algorithms and apply them
to the space based systems as
the high frequency, ground based detectors primarily look for short lived, burst like
signals, whereas the low frequency space based detectors primarily look for long lived, continuous
sources. One of the key differences between ground and space based gravitational wave detection
is ratio of the observation time to the orbital period of the detector. The ratio is
small for most high frequency sources, and of order one for most low frequency sources.
This means that LISA data analysis algorithms have to deal with orbital
signal modulation and multiple overlapping sources.

Here we describe a simple method for converting the output of a space based gravitational
wave detector, such as LISA, into useful astronomical information about the location and
physical characteristics a gravitational wave source. The method works by removing
the amplitude, phase and frequency modulation imparted by the orbital motion of the detector.
For monochromatic sources, the demodulation
procedure focuses the signal into a narrow frequency band, which creates a spike in
the demodulated power spectrum. By searching for sky locations that yield the largest
spikes it is possible to locate the brightest sources on the sky. These can then be subtracted
and the procedure iterated until all resolvable sources have been found. For
chirping sources the demodulation can be followed by a fast chirp transform~\cite{tom}, which
effectively re-concentrates the power that is spread by the frequency evolution of
the source. Once again the brightest sources on the sky can be identified and
iteratively subtracted to expose the weaker sources.

The demodulation procedure is remarkably simple. It utilizes the fact that space based
detectors such as LISA are able to return several interferometer outputs~\cite{aet}.
A general gravitational wave coming from a source in the ${\hat n}$ direction
can be expressed in barycentric coordinates as
\begin{equation}
h_{\mu\nu}(t,{\hat n}) = h_+(t,{\hat n}) e^+_{\mu\nu}({\hat n})
+ h_\times(t,{\hat n}) e^\times_{\mu\nu}({\hat n})
\end{equation}
where $h_+(t,{\hat n})$ and $h_\times(t,{\hat n})$ describe the two polarization states
with basis tensors $e^+_{\mu\nu}({\hat n})$ and $e^\times_{\mu\nu}({\hat n})$. The
polarization states $h_+$, $h_\times$ are related to the principal polarization states
$h_+^P$, $h_\times^P$ by the polarization angle $\psi$:
\begin{eqnarray}
&& h_+ = h_+^P\cos 2\psi + h_\times^P\sin 2\psi \nonumber \\
&& h_\times = h_\times^P \cos 2 \psi - h_+^P \sin 2 \psi \, .
\end{eqnarray}
In the idealized limit of a single
low frequency source and a noise-free detector, the 
interferometer outputs $s(t)$ can be expressed as
\begin{equation}\label{sig}
s(t) = {\bar h}_+(t,{\hat n}) D^+(t,{\hat n}) 
+ {\bar h}_\times(t,{\hat n})D^\times(t,{\hat n}) \,
\end{equation}
where $D^+(t,{\hat n})$ and $D^\times(t,{\hat n})$ are the sky-location
dependent interferometer response functions~\cite{cr1} and ${\bar h}_+(t,{\hat n})$ and
${\bar h}_\times(t,{\hat n})$
are the Doppler shifted gravitational wave strains
\begin{equation}\label{dop}
{\bar h}(t,{\hat n}) =  h\bigl(t + \frac{R}{c} \sin\theta\cos(2\pi f_m t -\phi),{\hat n}\bigr) \, .
\end{equation}
Here ${\hat n}\rightarrow (\theta,\phi)$ denotes the location of the source in ecliptic
coordinates, $R$ is
the Earth-Sun distance and $f_m=1/{\rm year}$ is the modulation frequency. A simple,
but impractical, approach for demodulating the signal is to form the combinations
\begin{equation}\label{n1}
{\bar h}_{\times}(t,{\hat n}) = \frac{s_{\rm I}(t)D_{\rm II}^+(t,{\hat n})
-s_{\rm II}(t)D_{\rm I}^+(t,{\hat n})}
{D_{\rm I}^\times(t,{\hat n})D_{\rm II}^+(t,{\hat n})-D_{\rm II}^\times(t,{\hat n})
D_{\rm I}^+(t,{\hat n})}
\end{equation}
and
\begin{equation}\label{n2}
{\bar h}_{+}(t,{\hat n}) = \frac{s_{\rm I}(t)D_{\rm II}^\times(t,{\hat n})
-s_{\rm II}(t)D_{\rm I}^\times(t,{\hat n})}
{D_{\rm I}^\times(t,{\hat n})D_{\rm II}^+(t,{\hat n})
-D_{\rm II}^\times(t,{\hat n})D_{\rm I}^+(t,{\hat n})}
\end{equation}
from interferometer outputs ${\rm I}$ and ${\rm II}$, then perform the coordinate transformation
\begin{equation}
t = t' - \frac{R}{c} \sin\theta\cos(2\pi f_m t' -\phi)
\end{equation}
to remove the leading order Doppler modulation~\cite{cl1}, thereby
recovering the demodulated strains ${\bar h}_+(t')$ and ${\bar h}_\times(t')$.
The main problem with this approach is that the denominators in Eqns.~(\ref{n1})
and (\ref{n2}) vanish at certain times of the year for sources within $60$
degrees of the ecliptic plane. When detector noise or other sources are present
the inversion becomes singular. Although this simple demodulation
procedure does not work in the time domain, it can be made to work in the frequency domain.

The key advantage of moving to the frequency domain is that the singularities in the temporal
inversion do not correspond to singularities in the frequency space inversion.
By Fourier transforming the data over a finite time interval, the demodulation procedure
can be re-cast as a linear algebra problem that can be solved using
the robust technique of singular value decomposition.

To simply our expressions we assume that the observation time $T$ is equal
to one year so that the frequency resolution $\Delta f =1/T$ is equal to the modulation
frequency $f_m$ (the general expressions are not hard to derive, but they
are more complicated). The Fourier space equivalents of
(\ref{sig}) and (\ref{dop}) become
\begin{equation}
s_k = A_{kl}^+({\hat n}) \, {\bar h}^+_l({\hat n}) + A_{kl}^\times({\hat n})\, 
{\bar h}^\times_l({\hat n}),
\end{equation}
and
\begin{equation}
{\bar h}^+_l = B_{ln}({\hat n}) \, h^+_n({\hat n}), \quad {\bar h}^\times_l = 
B_{ln}({\hat n}) \, h^\times_n({\hat n}),
\end{equation}
respectively. The matrices ${\bf A}$ and ${\bf B}$ are given by
\begin{equation}
A^+_{kl}({\hat n}) = D^+_{k-l}({\hat n}), \quad A^\times_{kl}({\hat n}) = D^\times_{k-l}({\hat n})
\end{equation}
and
\begin{equation}
B_{ln}({\hat n}) = J_{(l-n)}\bigl(2\pi n f_m \frac{R}{c}\sin\theta\bigr)
e^{i(l-n)(\pi/2-\phi)} \, .
\end{equation}
Here $J_q(x)$ is a Bessel function of the first kind of order $q$ and we have
used the Einstein summation convention for repeated indices. It is important to
note that the matrices ${\bf A}$ and ${\bf B}$ are both band diagonal as
$D^{+}_j({\hat n})\approx 0 \approx D^{\times}_j({\hat n})$
for $\vert j \vert > 4$ and $J_q(x)\approx 0$ for $\vert q \vert > x$. The next
step is to interleave the Fourier coefficients of the two polarizations $h^+(t)$ and
$h^\times(t)$ into a single column vector ${\bf h}$ such that $h_{2n}=h^+_n$ and
$h_{2n+1}=h^\times_n$. Similarly, the Fourier coefficients of the two interferometer
outputs $s_{\rm I}(t)$ and $s_{\rm II}(t)$ are interleaved into a single column vector ${\bf s}$.
The detector response can then be written as
\begin{equation}\label{fourier}
{\bf s} = {\bf M}({\hat n}) {\bf h}({\hat n}) \, ,
\end{equation}
where the modulation matrix ${\bf M}({\hat n})$ is the interleaved
product of ${\bf A}^+({\hat n})$,
${\bf A}^\times({\hat n})$ and ${\bf B}({\hat n})$:
\begin{eqnarray}
&& M_{2k\, 2l} = {A^+_{\rm I}}_{kj}B_{jl}, \quad
M_{2k\, 2l+1} = {A^\times_{\rm I}}_{kj}B_{jl}, \nonumber \\
&& M_{2k+1\, 2l} = {A^+_{\rm II}}_{kj}B_{jl}, \quad
M_{2k+1\, 2l+1} = {A^\times_{\rm II}}_{kj}B_{jl}.
\end{eqnarray}
If the interferometer outputs
are sampled $N$ times in a year, ${\bf M}({\hat n})$ will be a $2N\times 2N$
dimensional, complex, band diagonal matrix.
Multiplying ${\bf s}$ by the inverse of the modulation matrix
demodulates any sources located in the ${\hat n}$ direction. In practice, ${\bf s}$
will contain contributions from instrument noise ${\bf n}$, and multiple sources
${\bf h}^i({\hat n}^i)$:
\begin{equation}
{\bf s} = \sum_i {\bf M}^i({\hat n}^i) {\bf h}^i({\hat n}^i) + {\bf n} \, .
\end{equation}
We can estimate the contribution to the signal ${\bf s}$ from a source located
in the ${\hat n}$ direction by solving for the vector ${\bf h}^{\rm eff}({\hat n})$ that
satisfies the equation
\begin{equation}\label{svd}
{\bf s} = {\bf M}({\hat n}) {\bf h}^{\rm eff}({\hat n}) \, .
\end{equation}
The best fit solution is found using a singular value decomposition. Suppose that
source $j$ happens to lie in the ${\hat n}$ direction. Then
\begin{equation}
{\bf h}^{\rm eff}({\hat n}) \approx {\bf h}^j({\hat n}) + 
\sum_{i\neq j} {{\bf M}}^{-1}({\hat n}){\bf M}^i({\hat n}^i) {\bf h}^i({\hat n}^i)
+{{\bf M}}^{-1}({\hat n}){\bf n} \, .
\end{equation}
The key to the method is that sources at other sky locations are not demodulated, and
the noise is merely reshuffled, not amplified. Thus, we can use the
solution vector ${\bf h}^{\rm eff}({\hat n})$ as an estimate for the source
vector ${\bf h}^j({\hat n})$. Another key point is that the
modulation matrix is band diagonal, which allows us to perform the demodulation across
limited bandwidths using small modulation sub-matrices. If this were not the case we
would be confronted with the problem of inverting a $\sim 10^8 \times 10^8$ matrix.
Moreover, a source only has to compete with other sources that overlap with it in frequency space.

The demodulation procedure works for any type of gravitational wave source, but it is
especially useful when the source is a mildly eccentric, nearly Newtonian binary as the demodulated
signal is then approximately monochromatic. We can search for these sources by looking
for spikes in the the power spectrum of ${\bf h}^{\rm eff}({\hat n})$ for different sky
locations ${\hat n}$. A similar approach was employed in Ref.~\cite{cl1} using Doppler
demodulation of a single interferometer output.  Since the vast majority of LISA sources
are non-chirping, circular Newtonian binaries, the combined demodulation, power spike
search goes a long way toward solving the LISA data analysis problem.
After establishing the sky location and frequency of a source, the next step is to refine
the frequency measurement and extract the
Fourier amplitudes of the two polarizations $\tilde{h}^+$ and $\tilde{h}^\times$. Unless the
source frequency $f$ is an integer multiple of the sample frequency $\Delta f=f_m$,
the power will be shared by several discrete Fourier coefficients:
\begin{equation}
h^+_n \simeq \tilde{h}^+ {\rm sinc}(\pi x_n) e^{i \pi x_n} \quad {\rm where} \quad x_n=f/f_m -n,
\end{equation}
and similarly for $h^\times_n$.
%One way to pinpoint the frequency is
%to minimize the quantity
%\begin{equation}\label{chi}
%\chi^2 = \sum_{j} \left(\frac{\vert h^+_{q+j}\vert}{\vert h^+_q \vert} - 
%\frac{\vert \delta \vert}{\vert \delta - j\vert}\right)^2 +
%\left(\frac{\vert h^\times_{q+j}\vert}{\vert h^\times_q \vert} - 
%\frac{\vert \delta \vert}{\vert \delta - j\vert}\right)^2,
%\end{equation}
%where $q$ is the nearest integer to $f/f_m$ and $\delta =f/f_m-q \in (-1/2,1/2]$ is
%the remainder. Using the best-fit value of $f$ from (\ref{chi}), the Fourier
%amplitude $\tilde{h}^+$ can be determined by minimizing the quantity
%\begin{equation}\label{chiplus}
%\chi_+^2 = \sum_n \vert h^+_n - \tilde{h}^+ {\rm sinc}(\pi x_n) e^{i \pi x_n} \vert^2
%\end{equation}
%and similarly for $\tilde{h}^\times$. (The choice of norms in equations (\ref{chi}) and
%(\ref{chiplus}) may not be optimal).
Now suppose that the integration period $T$ is changed slightly to $T'=T(1+\epsilon)$ where
$\epsilon \ll 1$. The Fourier coefficients of $h(t)$ evaluated over the interval $T'$
are related to those evaluated over the interval $T$ by
\begin{equation}
h'_n = \sum_j h_j {\rm sinc}(\pi x_{jn}) e^{i \pi x_{jn}} ,
\end{equation}
where
\begin{equation}
x_{jn} = j\left(\frac{T'}{T}\right) - n \, .
\end{equation}
If the source frequency $f$ is an integer multiple of the new sample frequency
$\Delta f' = 1/T'$, all the power will be concentrated in a single Fourier mode
$h'_k$. Thus, the frequency spreading can be removed by finding the value of $\epsilon$
that maximizes the spike in the stretched power spectra of $h^+(t)$ and $h^\times(t)$.
This procedure yields an improved estimate of the source frequency, and a direct
estimate of the Fourier amplitudes of the gravitational wave:
\begin{equation}
\tilde{h}^+ \simeq {h^{+}}'_{k}e^{i \delta} \quad {\rm and}
\quad  \tilde{h}^\times \simeq {h^{\times}}'_{k}e^{i \delta}.
\end{equation}
The quantity $\delta = 2 \pi f(T'-T)$ accounts for the phase rotation imparted by
the change in integration period.
Once the Fourier amplitudes of the two polarizations
have been estimated they can be used to calculate the amplitude ${\cal A}$,
inclination $\iota$, polarization angle $\psi$ and orbital phase
$\varphi_0$ of the binary using the relations
\begin{equation}
\psi= \frac{1}{4}\left(\arctan \left(\frac{\tilde{h}^I_+ 
+ \tilde{h}_\times^R}{\tilde{h}_\times^I -
\tilde{h}_+^R}\right) + \arctan\left(\frac{ \tilde{h}_+^I - \tilde{h}_\times^R}
{\tilde{h}_+^R + \tilde{h}_\times^I}\right) \right),
\end{equation}
\begin{equation}
\varphi_0= \frac{1}{2}\left(\arctan\left(\frac{ \tilde{h}_+^I - \tilde{h}_\times^R}
{\tilde{h}_+^R + \tilde{h}_\times^I}\right) - \arctan \left(\frac{\tilde{h}^I_+ 
+ \tilde{h}_\times^R}{\tilde{h}_\times^I -
\tilde{h}_+^R}\right)\right),
\end{equation}
\begin{equation}
{\cal A} = \frac{A_+ + \sqrt{A_+^2 - A_\times^2}}{2}
\end{equation}
\begin{equation}
\iota = \arccos\left(\frac{-A_\times}{A_+ + \sqrt{A_+^2 - A_\times^2}}\right) \, ,
\end{equation}
where
\begin{eqnarray}
&& A_+ = \frac{4(\cos 2\psi\, \tilde{h}_+^R - \sin 2\psi\, \tilde{h}_\times^R)}{\cos\varphi_0}
\nonumber \\ \nonumber \\
&& A_\times = \frac{4(\sin 2\psi \,  \tilde{h}_+^R + \cos 2 \psi \,
\tilde{h}_\times^R)}{\sin\varphi_0} \, ,
\end{eqnarray}
and $\tilde{h}_+^R = \Re ( \tilde{h}^+)$, $\tilde{h}_+^I = \Im ( \tilde{h}^+)$,
$\tilde{h}_\times^R = \Re ( \tilde{h}^\times)$, and
$\tilde{h}_\times^I = \Im ( \tilde{h}^\times)$.

%\begin{eqnarray}\label{wf}
%\tilde{h}_+^R &=& \frac{1}{4}\left(A_+\cos 2\psi \cos\varphi_0 +A_\times \sin 2\psi
%\sin\varphi_0\right) \nonumber \\
%\tilde{h}_+^I &=& \frac{1}{4}\left(A_+\cos 2\psi \sin\varphi_0 -A_\times \sin 2\psi
%\cos\varphi_0\right) \nonumber \\
%\tilde{h}_\times^R &=& \frac{1}{4}\left(A_\times \cos 2\psi
%\sin\varphi_0 - A_+\sin 2\psi \cos\varphi_0 \right) \nonumber \\
%\tilde{h}_\times^I &=& -\frac{1}{4}\left(A_\times\cos 2\psi
%\cos\varphi_0 + A_+\sin 2\psi \sin\varphi_0 \right) ,
%\end{eqnarray}
%where
%\begin{equation}
%A_+ = {\cal A}(1+\cos^2 \imath), \quad A_\times =  2{\cal A}\cos \imath \, .
%\end{equation}
%The gravitational wave frequency $f$ and overall amplitude ${\cal A}$ are related to the
%masses of the stars $M_1, \, M_2$, their orbital separation $r$, and the distance from the
%detector $D$, by
%\begin{equation}
%{\cal A} = \frac{2 M_1 M_2}{r D} \quad {\rm and} \quad
%f = \frac{(M_1+M_2)^{1/2}}{\pi r^{3/2}} \, .
%\end{equation}

\begin{figure}[t]
\vspace{95mm}
\includegraphics{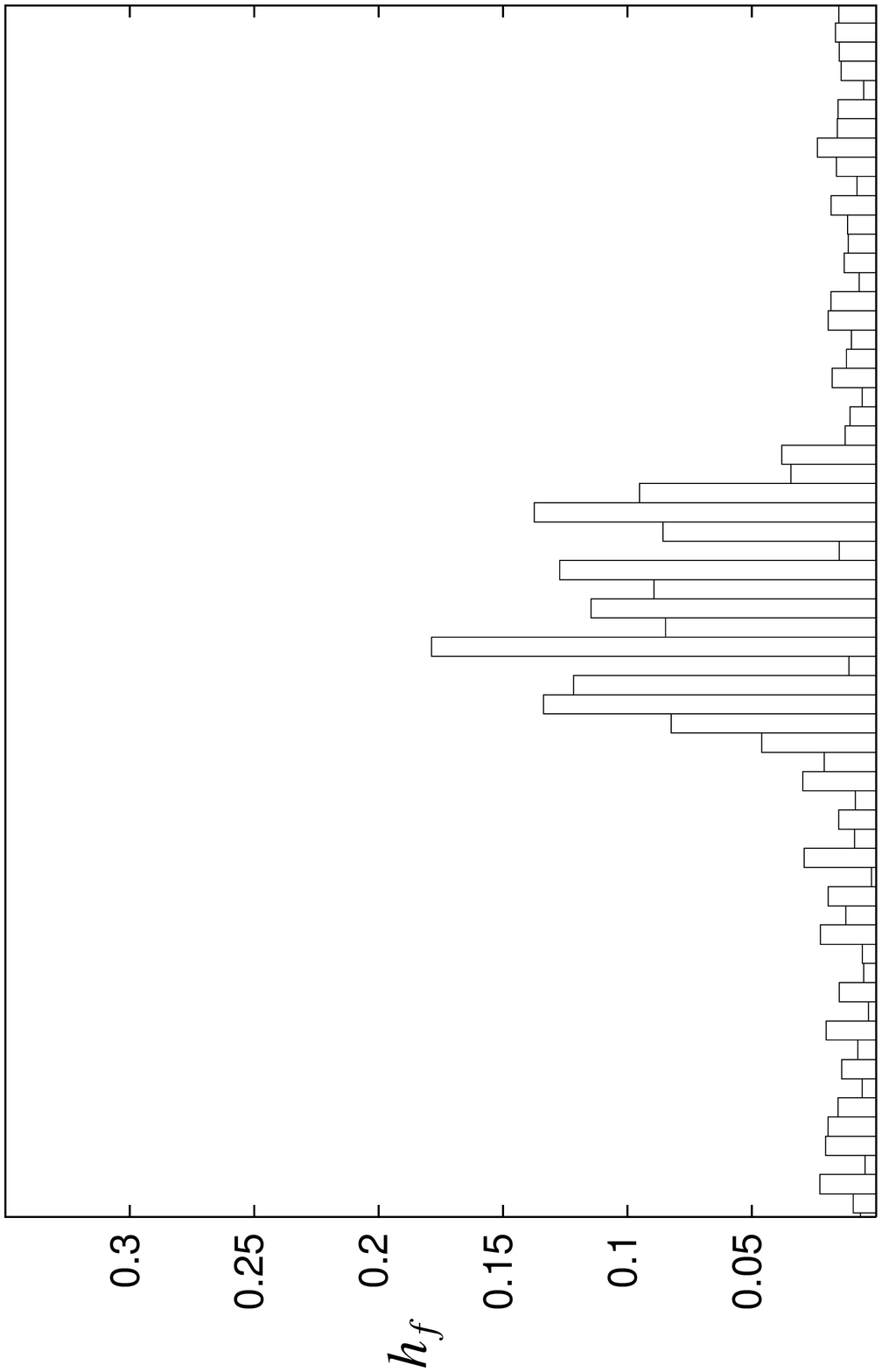}
\includegraphics{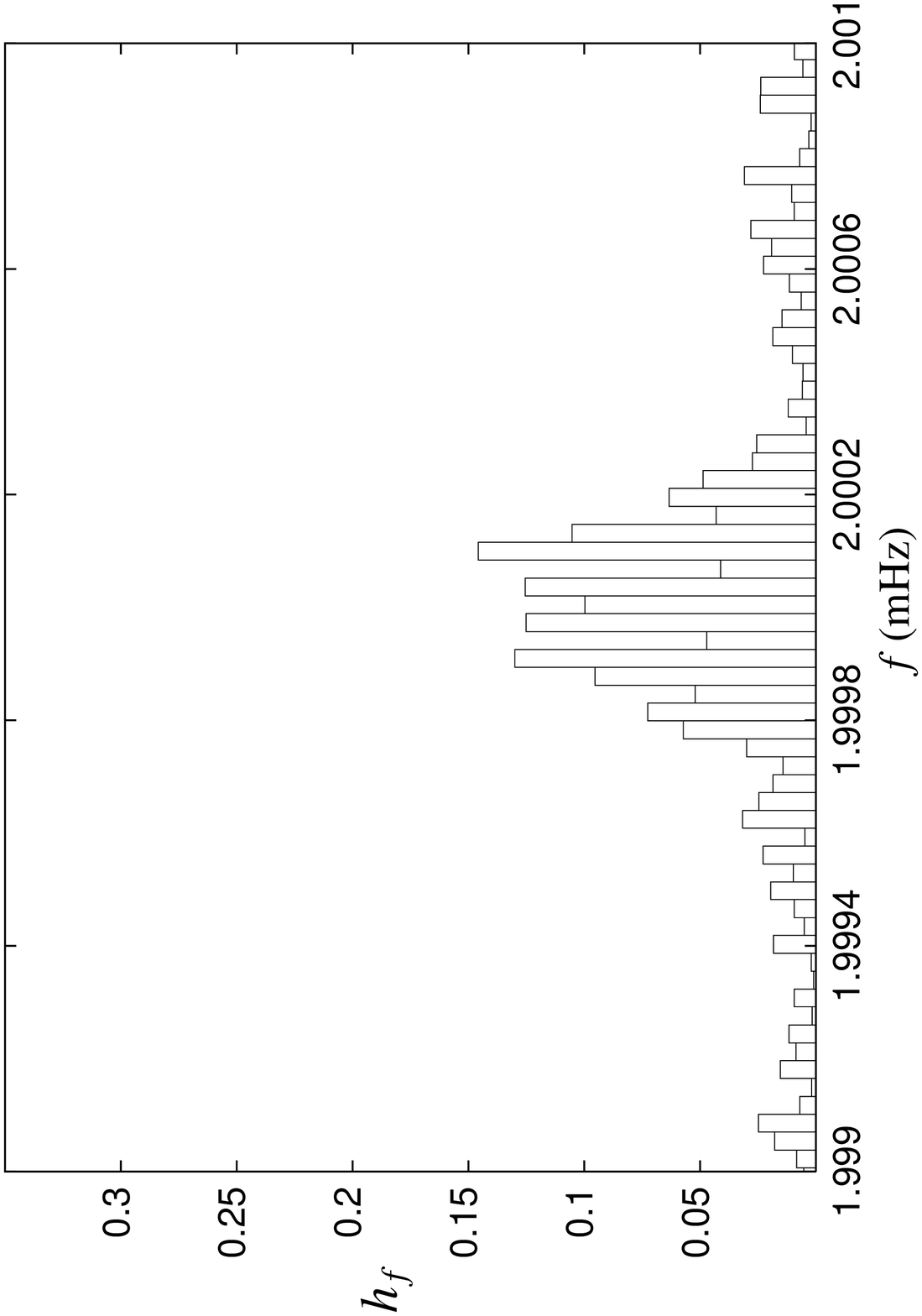}
\vspace{5mm}
\caption{The RMS strain spectral density in LISA channels $s_{\rm I}$ (upper panel) and $s_{\rm II}$
(lower panel) in units of $10^{-22}$.}
\end{figure}

As a typical example of a LISA source, consider a White Dwarf binary with
orbital period $1000$ seconds, comprised
of two $0.5 M_{\odot}$ White Dwarfs, located 2 Kpc from Earth in the direction of the galactic center
$(\theta, \phi) = (95.54^\circ, 266.83^\circ)$. Setting $\psi = 111.2^\circ$, $\iota = 60^\circ$
and $\varphi_0 = 20.1^\circ$, and using the standard LISA noise curve, we arrive at the signal
strengths shown in Fig.~1. The raw signal to noise ${\rm SNR}_{r}$, and the optimal signal
to noise that can be achieved by matched filtering, ${\rm SNR}_{m}$, are defined:
\begin{eqnarray}
&& {\rm SNR}_{r} = \left(\frac{\int_{\rm bw} S_h \, df}{\int_{\rm bw} S_n \, df}\right)^{1/2} \nonumber \\
&& {\rm SNR}_{m} = \left(\int_{\rm bw} \frac{2T \, S_h \, df}{S_n }\right)^{1/2}\, , \nonumber 
\end{eqnarray}
where $S_h$ is the power spectral density of the signal and $S_n$ is the power spectral density
of the noise, and the integration is across the bandwidth of the signal. Using these definitions
we find that ${\rm SNR}_{r}^{\rm I} = 5.4$, ${\rm SNR}_{r}^{\rm II} = 5.2$, and the optimal
signal to noise is ${\rm SNR}_{m} = 49.9$. Applying the demodulation procedure yields the two
gravitational wave polarizations shown in Fig.~2. The source parameters are recovered to an
accuracy of $\Delta f = 0.008 f_m$, $\Delta \theta = 1.4^\circ$, $\Delta \phi = 0.2^\circ$,
$\Delta {\cal A}/{\cal A} = 0.054$, $\Delta \iota = 0.04^\circ$, $\Delta \psi = 16.4^\circ$
and $\Delta \varphi_0 = 8.6^\circ$. These uncertainties are comparable to those found
using matched filtering, so the demodulation method is not only extremely fast, but
also close to optimal.

\begin{figure}[t]
\vspace{95mm}
\includegraphics{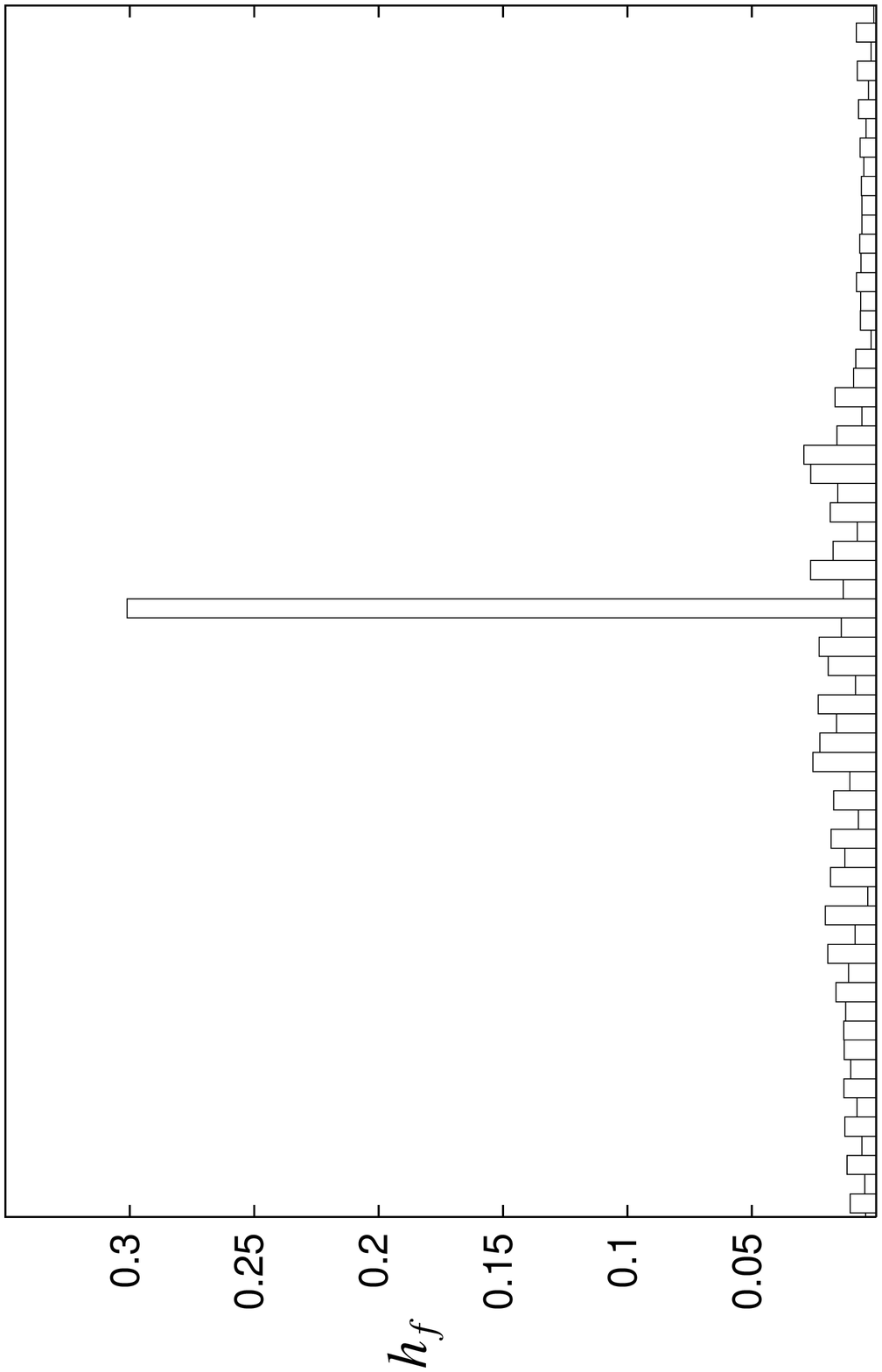}
\includegraphics{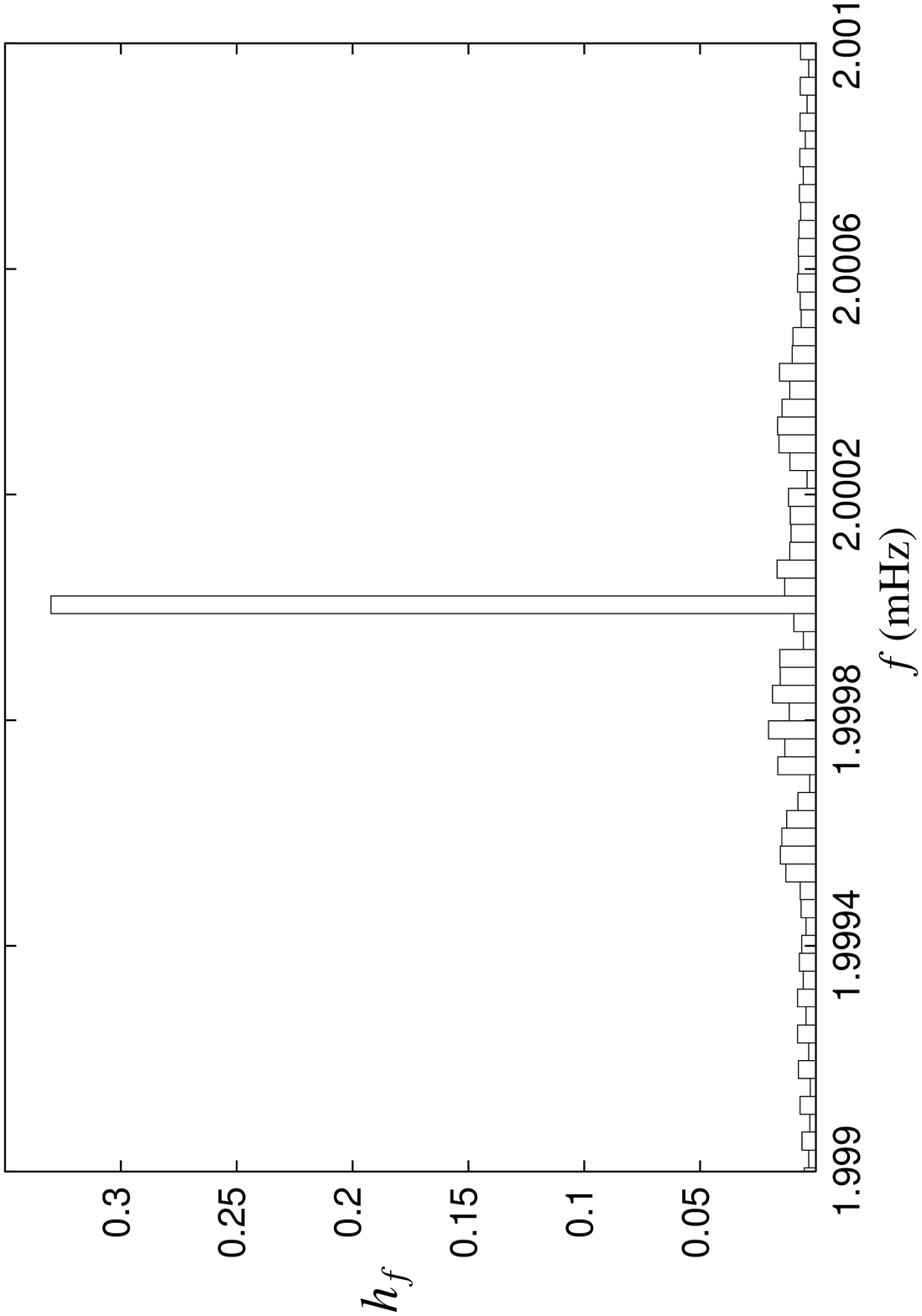}
\vspace{5mm}
\caption{The RMS strain spectral density of the demodulated plus $\tilde{h}_+$ (upper panel) and
cross $\tilde{h}_\times$ polarizations (lower panel) in units of $10^{-22}$.}
\end{figure}

The method continues to perform well when the SNR ratios are smaller, or when there are
multiple sources in the same frequency band. Even with multiple sources the procedure
delivers a good initial fit for the parameters describing each source. By
combining the demodulation procedure with the {\em gCLEAN} algorithm described in Ref.~\cite{cl2},
it will be possible to extract thousands of individual sources from the galactic background
at a modest computational cost.

The simple power spike search will not work for rapidly evolving systems, such as supermassive black hole
binaries, as even the demodulated power spectrum will be spread over thousands of frequency
bins. If one thinks of a Fourier transform as a matched filter using sines and cosines, we see
that we get a good match for monochromatic sources (the power spike), but a poor match for
chirping sources. The solution is to employ a different kind of transform that uses filter
functions that evolve in amplitude and frequency -  the Fast Chirp Transform~\cite{tom}. Just as
a Fast Fourier Transform (FFT) allows us to detect signals with a constant frequency, the Fast
Chirp Transform (FCT) allows us to detect signals with variable frequency. By applying a FCT rather
than a FFT to the demodulated signal, we can search for the sky locations that harbour the
brightest chirping sources and extract their physical characteristics.

Much remains to be done in the field of LISA data analysis, but fast non-template based methods
such as the total demodulation procedure and the Fast Chirp Transform will likely play a key
role in future developments.

The work reported here
is supported by the NASA EPSCoR program through Cooperative Agreement NCC5-579.

\end{document}